
\input phyzzx

\REF\hawi{S.~W.~ Hawking
\journal Comm .Math. Phys. & 43 (75) 199.}

\REFS\kith{~W.~ Kim, K.~S.~ Thorne
\journal Phys. Rev. & D43 (91) 3929.}

\REF\hawii{S.~W.~ Hawking
\journal Phys. Rev. & D46 (92) 603.}

\REF\zeld{Ya.~B.~ Zel'dvich
\journal Pis'ma Zh. Eksp. Teor. Fiz. &12 (70) 443.
;{\sl JETP~ Lett. \bf 12 }(1988) 307.}

\REF\bida{N.D. Birrel and P.C.W. Davies
{\it Quantum Fields in Curved Space},  Cambridge Univ. Press, (1982)
 and references therein.}

\REF\nato{H.~ Nariai, K.~ Tomita
{\it General Relativistic Cosmology}; Chap.~9 (in Japanese),
Shokabo, (1988)}

\REF\gupi{A.~ Guth, So-Y.~ Pi
\journal Phys. Rev. Lett. & 49 (82) 1110.}

\REF\hawiii{S.~W.~ Hawking
\journal Phys. Lett. & B115 (82) 295.}

\REF\star{A.~ Starobinsky
\journal Phys. Lett. & B117 (82) 175.}

\REF\cghs{C.~G.~Callan, S.~B.~Giddings, J.~A.~Harvey and
A.~Strominger
\journal Phys. Rev. & D45 (92) R1005.}

\REF\polya{A.~M.~ Polyakov
\journal Phys. Lett. &B103 (81) 207.}

\REF\hs{J.~A.~Harvey, A.~Strominger {\it Quantum Aspects of Black Hole},
EFI-92-41,
 September 1992 and references therein.}

\REF\mr{D. F. Mazzitelli, J. G. Russo, {\it Dilaton Quantum Cosmology
in Two-
Dimensions}, UTTG-28-92, November 1992.}

\REF\mn{T. Mishima, A. Nakamichi, {\it Chronology Protection in
Two-Dimensional Dilaton Gravity}, TIT/HEP-218/COSMO-29, April 1993.}

\REF\yos{M.~Yoshimura, {\it Quantum Decay of de Sitter Universe and
Inflation without Fine Tuning in 1 + 1 Dimensions},
TU/92/416, September 1992, \nextline
M.~Hotta, Y.~Suzuki, Y.~Tamiya and M.~Yoshimura, {\it Quantum Back
Reaction in Integrable Cosmology},
TU/93/426, January 1993, \nextline
M.~Hotta, Y.~Suzuki, Y.~Tamiya and M.~Yoshimura, {\it Universality
of Final State in Two-Dimensional Dilaton Cosmology},
TU/93/431, April 1993.}

\REF\rsti{J.~G.~Russo, L.~Susskind, L.~Thorlacius,
\journal Phys. Rev. & D46 (92) 3444.}

\REF\hael{S.~W.~ Hawking, G.~F.~R.~ Ellis, {\it The Large Scale
Structure of Space-Time}, Cambridge Univ. Press, (1973).}

\REF\soda{J.~ Soda, {\it Hierarchiral Dimensional Reduction and Gluing
Geometry}, (1992)}

\REF\mntu{Y.~ Matsumura, N.~ Sakai, Y.~ Tanii, T.~ Uchino,{\it Correlation
Functions in Two Dimensional Dilaton Gravity},
TIT/HEP-204, STUPP-92-131, August 1992.}

\REF\kaz{S. Hirano, Y. Kazama, Y. Satoh, {\it Exact Operator Quantization  of a
Model of Two-Dimensional Dilaton Gravity},
UT-Komaba 93-3, March 1993.}

\def\p{\partial}
\def\phip{\p_+\phi}
\def\phim{\p_-\phi}
\def\phipm{\p_+\p_-\phi}

\def\rhopm{\p_+\p_-\rho}

\def\xp{x^+}
\def\xm{x^-}
\def\up{u_+}
\def\um{u_-}
\def\xchon{x^{\prime}}


\pubnum={225/COSMO$-$33}
\nopagenumbers
\titlepage
\vskip 0.5cm
\title{
Cosmological Model in 2d Dilaton Gravity
\footnote{\diamond}
{An expanded version of the talk presented by T. Mishima
at Yukawa Institute of Theoretical Physics workshop `Quantum Gravity'
(24-27, November 1992).}
 }
\vskip 1cm
\author{Takashi Mishima \footnote{\dagger}{On
leave of absence from Institute for
Nuclear study, University of Tokyo, Midori-cho, Tanashi-shi, Tokyo 188,
Japan}
      \ and \ Akika Nakamichi \footnote{\ast}{e-mail address:
       akika@phys.titech.ac.jp }
        }
\vskip 0.5cm
\address{
  ${\dagger}$ Physical Science Laboratories, College of Science and
Technology,
    \nextline Nihon University, Narashinodai, Funabashi-shi, Chiba 274,
Japan
 \nextline
 \nextline
  ${\ast}$ Department of Physics,
    Tokyo Institute of Technology
    \nextline Oh-okayama, Meguro-ku, Tokyo 152, Japan
        }

\vskip 0.5cm

\abstract{We apply CGHS-type dilaton gravity model to
(1+1)-dimensional cosmological situations. First the behavior of a compact
1-dimensional universe (i.e. like a closed string) is classified on the
assumption of homogeneity of universe. Several interesting solutions are found,
which
include a Misner-type universe having
closed time-like curves, and an asymptotically de Sitter universe first pointed
out by Yoshimura. In the second half
of this talk, we discuss the modification of the classical homogeneous
solutions, considering inhomogeneity of classical conformal matters and also
quantum back-reaction respectively.}

\vfill
\eject


\chapter{Introduction}


	In order to understand  the evolution of the very early universe or
the physics of a  very small black hole, one must consider the quantum effects
affected by a strong gravitational field and further quantum mechanical
treatment of gravitational degrees of freedom itself.
So far we have not yet succeeded in the establishment of
a quantum theory of gravity, although some unified theory
may describe the short-distance physics in Planck scale.
Hence several theoretical physicists have used the semi-classical approach
as a well formulated first step,
that is, the gravity is treated classically and
matter fields are treated quantum mechanically.
Using this approach, there have been a lot of studies on the
quantum physics in black hole spacetime
or in other interesting space-time ;
 (a) Hawking radiation  [\hawi],
 (b) Quantum instability of space-time having closed time-like curves
      [\kith, \hawii],
 (c) Decreasing of inhomogeneity due to the particle creation
       [\zeld, \bida],
 (d) Avoiding an initial singularity which is expected from the
         singularity theorem   [\nato],
 and
 (e) Quantum generation of density fluctuations in inflationary era
     [\gupi, \hawiii, \star]
(See [\bida] for general references).
However in real 3+1 dimensional spacetime,
even by using semi-classical approach we are forced a very
formidable task.
Then to proceed the analysis, another approximation was introduced generally.
 Indeed in the cases (a), (b) and (e), quantum back-reaction on the
background geometry was neglected generally.
 In the cases (c) and (d),  gravitational degrees of freedom
 were reduced to be finite by restricting the form of metric.
 In 3+1 dimensions, the back-reaction problems have not yet been  solved
in closed form.

	Recently a manageable and interesting
1+1 dimensional toy model was proposed by Callan, Giddings, Harvey
and Strominger [\cghs] ( abbreviated as CGHS).
They attempted to analyze the back-reaction of the Hawking radiation[\hawi]
on the two-dimensional analogue of black hole geometry
in a consistent way by the use of this model.
In their first paper, CGHS developed their original scenario as follows.
First, introducing 1+1 dimensional dilaton gravity,
CGHS found the solutions corresponding to black hole formation,
and showed occurrence of the Hawking radiation
in the spacetime-background they obtained.
Next taking account of the quantum effect of quantized conformal matters
as the Polyakov term[\polya], they analyzed the back-reaction of
the Hawking effect in a leading semi-classical approximation,
however their first scenario on the quantum black hole
remains to be elaborated. Up to the present time,
many people have studied the behavior of
a quantum black hole in this model[\hs].

	On the other hand, in general relativity, there are other interesting
problems which must be resolved taking account of the quantum effects than
the problem of Hawking radiation.  In the present paper,
we apply this model to compact 1+1 dimensional cosmology.
(See also [\mr, \mn])
First, after brief review of the model in section 2,
the behavior of classical homogeneous solutions are clarified
in sections 3 and 4. Several interesting solutions are found,
including 1+1 dimensional analogue of the Misner's universe
which has closed time-like curves (CTC's),
and the universe having asymptotic de Sitter space region
first discovered by Yoshimura [\yos].
In the second half of the paper,
the modifications of the behavior clarified in the previous sections are
considered, when we introduce inhomogeneous classical conformal matter and
the quantum effect respectively. Here we especially give attention to
whether the laws of physics in the CGHS model prevent the appearance of
closed time-like curves or not,
following the `Chronology Protection conjecture' suggested
by Hawking [\hawii].
In section 5, the influence of inhomogeneous classical matter fields
on the basic homogeneous behaviour of the universe is considered.
In section 6 and 7, we consider the quantum back-reaction effect,
using the simple model first introduced by Russo, Susskind and Thorlacius
[\rsti].
\chapter{Classical field equations}


 We consider the $ 1+1 $ dimensional renormalizable theory of gravity
 coupled to a dilaton scalar field $\phi$ and $N$ massless conformal
 fields $ f_i$ [\cghs].
 The classical action is
$$
\eqalign{
  S
 = {1 \over 2\pi} \int d^2 x \sqrt{-g}
   \bigl[e^{-2\phi} (R + 4(\nabla \phi)^2 - 4\lambda^2)
   -  {1 \over 2} \sum^N_{i=1} (\nabla f_i)^2 \bigr]                   \cr
        } \, ,    \eqno\eq
$$
where $R$ is the scalar curvature, $\lambda^2$ is a cosmological
 constant.
 In this paper we treat the model with negative cosmological term,
which is different from the original CGHS model in the sign of the
cosmological term.
 We also consider the case of positive sign of the matter term,
first introduced by Yoshimura[\yos].
 We call the former case type-I, and the latter case type-II.
 Mainly we proceed with the analysis using type-I as an example.

The equations of motion derived from (2.1) are

$$
\eqalign{
  4e^{-2\phi} \bigl[ \nabla_\mu \nabla_\nu \phi
   - g_{\mu\nu} \nabla^2 \phi + g_{\mu\nu} (\nabla \phi)^2
   + g_{\mu\nu} \lambda^2 \bigr]
  =& \sum_{i=1}^N \bigl( \nabla _\mu f_i \nabla _\nu f_i
   -{1\over 2} g_{\mu\nu} (\nabla f_i)^2 \bigr) \, ,                 \cr
R
 =&  4(\nabla \phi)^2 - 4\nabla^2 \phi + 4\lambda^2 \, ,             \cr
 \nabla^2 f_i
 =& \> 0 \, .                                                        \cr
         }
\eqno\eq
$$

The conformal gauge is adopted:
$$
 g_{\mu\nu} dx^{\mu} dx^{\nu}
 = -e^{2\rho} dx^{+} dx^{-} \, ,
\eqno\eq
$$
where $x^{\pm} = t \pm x$ . The equations of motion then reduce to
$$
\eqalign{
 0
 =& - 4\phipm + 2\rhopm + 4\phip \phim - \lambda^2 e^{2\rho}   \,,     \cr
 0
 =& \phipm - \rhopm \,,                                             \cr
 0
 =& \p_{+}\p_{-} f_{i} \,.                                          \cr
        }
\eqno\eq
$$

 In addition, the following constraints have to be imposed:
$$
4e^{-2\phi}(\p_{\pm}^{2}\phi -2\p_{\pm} \rho \p_{\pm} \phi)
 = \sum_{i=1}^N \p_{\pm}f_{i} \p_{\pm}f_{i} \, .
\eqno\eq
$$

We adopt the periodic boundary condition that the spacetime point $(t,x)$
is identified with $(t,x+L)$ and the initial
condition that the universe starts from a static cylinder spacetime
endowed
with usual Minkowski metric at the past infinity.
Then the general form of the solutions is given by:
$$
\eqalign{
e^{-2\phi}
 =& \up + \um + e^{-2\lambda t}   \, ,               \cr        e^{2\rho}
 =& e^{-2\lambda t}e^{2\phi}   \, ,                                   \cr
         }
\eqno\eq
$$
where $u_+$ and $u_-$ are chiral periodic functions which satisfy the
following equations from the constraints (2.5):
$$
 0
 = \p _\pm ^2 u_\pm + \lambda \p _\pm u_\pm
   + {1 \over 2} \p _\pm f \p _\pm f \, .
\eqno\eq
$$

\chapter{Vacuum solutions}
 If there is no matter field, general solutions that satisfying the
periodic boundary condition depend only on time:
$$
\eqalign{
 e^{2\phi}
 =& (M + e^{-2\lambda t})^{-1} \, ,                                   \cr
 e^{2\rho}
 =& e^{-2\lambda t}e^{2\phi}   \, ,                                   \cr
         }
\eqno\eq
$$
where $M$ is an arbitrary constant. We classify the behavior of the
solutions
into three types with respect to the sign of  $M$ (Fig.1).

 When $M$ equals to 0, the solution becomes an analogue of the Linear
Dilaton
Vacuum solution in CGHS model.  The world is a static
cylinder spacetime as shown in Fig.1(a).

  When $M$ is negative (Fig.1(b)),
we see from (3.1) that the observer meet some
singularity in a finite proper time. From the expression of scalar curvature
$R$:
$$
 R
 = -{ 4\lambda^2 M \over M+e^{-2\lambda t} } \, ,
\eqno\eq
$$
we can see that this singularity is a true singularity.
In fact this singularity is the same as the one in the 1 + 1 dimensional
black-hole treated in CGHS.

  On the other hand when $M$ is positive (Fig.1(c)),
the space collapses into zero volume
in a finite proper time(as coordinate time $t$ goes to $+\infty$).
However from (3.2) the scalar curvature still remains a finite value
$-4\lambda^2$
at the point. Hence we expect that the spacetime can be extended.
In fact, if one defines the coordinates as:
$$
\left \{ \matrix{
 \eta
 =& -e^{-2\lambda t}  \, ,                        \cr
 \psi
 =& t \pm x           \, ,                        \cr
                }
\right.
\eqno\eq
$$
the metric becomes
$$
 ds^2
 = {-1 \over M - \eta} (\eta d\psi^2 + {1 \over \lambda} d\psi d\eta) ,
\eqno\eq
$$
which is analytic in the extended manifold defined by
$\psi$ and by $-\infty < \eta < M$.

The behavior of the extended manifold is shown in Fig.2.
The region $\eta < 0$ is isometric with the previous manifold.
The region $\eta > 0$ is extended part, where the closed timelike curves
appear, because the roles of $t$ and $x$ are interchanged.
The surface $\eta$ = 0 is the boundary of the Cauchy Development;
that is the Cauchy Horizon.
 This extension is achieved by the same way as in the case of Misner space
(we have two dimensional version when $\lambda  = 0 $) and Taub-NUT space
[\hael]. Hawking used the Misner space to discuss the chronology protection
conjecture [\hawii].
\chapter{Solutions with homogeneous matter}

The homogeneous solution of the massless scalar field equation
is as follows,
$$
 f
 = \sqrt{2\varepsilon}\, t
\eqno\eq
$$
where the constant $\varepsilon$ means the energy density of matter.
In type-I case, the general solution with homogeneous matter is
the following (discussed also by Mazzitelli and Russo [\mr]),
$$
\eqalign{
 e^{2\phi}
 =& (M - {\varepsilon \over 2\lambda}t + e^{-2\lambda t})^{-1} \, ,    \cr
 e^{2\rho}
 =& e^{-2\lambda t}e^{2\phi}   \, .                                   \cr
         }
\eqno\eq
$$
 The solution becomes singular in finite $t$, due to the
positivity of energy density $\varepsilon$. This singularity is real one,
because the curvature becomes infinite.
In contrast to the case in four dimensions, this singularity means an infinite
expansion
of the space and infinite divergence of
`gravitational constant' $e^{2\phi}$.
 Such behaviour is not so unusual from the viewpoint that the CGHS model
is
similar to two dimensional models induced from
spherical symmetric or plane-symmetric four dimensional gravity [\soda].
 For example,
the behavior stated above in the CGHS model corresponds
to the behavior of the radial coordinate and
the inverse of `proper radius' in the case of spherical symmetric collapse.
 Same correspondence is also seen between the CGHS model
without cosmological term and the Bianchi-type I homogeneous
universe model.

 Next we consider the model of type-II which differs from the previous model
(type-I) in the sign of the matter term.
In this model we change the sign of the energy density
$\varepsilon$ in (4.2).
The behavior of the universe is classified according to
the size of $\varepsilon$.

When $\varepsilon$ is sufficiently small, the universe meets with a future
singularity in $t= +\infty$ (in the finite proper time).
This singularity is different from type-I.
The space collapses into 0, where the curvature becomes infinite.

 When $\varepsilon$ is sufficiently large, the gravitational constant
$e^{2\phi}$ approaches to $+\infty$ in the
finite time, where the
evolution of the universe comes to the end.
This singularity is the same one as in the case of type-I.

Finally it should be pointed out that the model of type-II has an
interesting behavior found by Yoshimura [\yos],
if the energy density $\varepsilon$ is adjusted
so that the solutions of
$$
\eqalign{
 0
 =& e^{-2\phi}                                             \cr
 =&  M + {\varepsilon \over 2\lambda} + e^{-2\lambda t}    \, ,
        }
\eqno\eq
$$
are degenerate with respect to $t$. Here the degenerate solution is denote by
`$t_0$'.
Then we expand $e^{2\rho}$ in $s=\, t-t_0$ as follows,
$$
\eqalign{
 e^{2-\phi}
 =& 2\lambda ^2 s^2 e^{-2\lambda t_0} + O(\, s^3 \,) \, ,          \cr
 e^{2\rho}
 =& {1 \over 2\lambda^2 s^2} +  O(\, {1 \over s} \,) \, .                  \cr
        }
\eqno\eq
$$
{}From eq.(4.4), it is easily shown that the universe asymptotically approaches
to de Sitter spacetime.

\chapter{Modifications due to classical inhomogeneity  }

In this section, we consider modifications of the basic homogeneous
behavior of the universe stated in the previous sections,
when inhomogeneous classical conformal
matters are introduced.
The solutions of (2.4) satisfying the periodic boundary condition
always exist for an arbitrary configuration of the scalar fields.
We expand the scalar fields in Fourier series:
$$
\eqalign{
 f(\xp, \xm)
 =& f_+ (\xp) + f_- (\xm) \, ,              \cr
 f_{\pm}
 =& \alpha_{\pm} + \sqrt{{\varepsilon \over 2}} x^{\pm}
    + \sum^{\infty} _{n=1}\bigl( a^{\pm}_n \sin {2\pi n \over L} x^{\pm}
               + b^{\pm}_n \cos {2\pi n \over L} x^{\pm} \bigr) \, , \cr
        }
\eqno\eq
$$
where $\alpha_{\pm} , a^{\pm}_{n}$ and $b^{\pm}_{n}$ are the expansion
coefficients.
Then the solution is as follows,
$$
\eqalign{
 e^{-2\phi}
 =& M
   \mp \bigl\{ {\varepsilon \over 2\lambda}
       + {1 \over 2\lambda} \Sigma \bigr\} t
   + e^{-2\lambda t}
   + \bigl(\, oscillation\  part \, \bigr)\, ,    \cr
 \Sigma
 =& \sum_{(\cdot)=\pm} \sum^{\infty} _{n=1}\,({2\pi n \over L})^2 \,
[(a^{(\cdot)}_{n}) ^{2}
                                     + (b^{(\cdot)}_n) ^2]\, ,          \cr
        }
\eqno\eq
$$
where the fourth term on the right hand side of the first equation in (5.2)
is the oscillation part, that is the sum of the trigonometric functions.
The models of type-I and of type-II correspond to negative sign and
positive sign in front of the linear-$t$ term in (5.2) respectively.
{}From (5.2) it should be noted that any classical configuration
of matter fields makes a finite contribution to the term proportional to time,
which causes divergence of the scalar curvature. Therefore the universe
inevitably meets singularity at $t=\infty$ and cannot extend to the region
with closed time-like curves both in type-I and in type-II.

Next we investigate how the oscillation part contributes to the global
structure. Especially we give attention to changes of the property of
the singularity and the possibility of local exponential-expansion
of the universe by introducing the
inhomogeneity. As a simple example, a standing wave solution is considered:
$$
  f_{\pm}
 = a\; \sin {2\pi n \over L} x^{\pm} \, ,
\eqno\eq
$$

In the case of type-I, the corresponding solution of the dilaton field is:
$$
\eqalign{
 e^{-2\phi}
 =& M
   -  {a^2 \omega_n ^2 \over 2\lambda}\, t
   + e^{-2\lambda t}
   - {a^2 \omega_n /\,4 \over \sqrt{\lambda^2 + 4\omega_n ^2}}
    \sin(2\omega_n t + \theta)\, \cos(2\omega_n x) \, ,           \cr
 \tan (\theta)
 =& -{2\omega_n \over \lambda} \, .                               \cr
        }
\eqno\eq
$$
We now try to find which positions $x$ give the degenerate solutions for
the equation: $e^{-2\phi} = 0$ with respect to $t$.
Following the discussion in the last part of section 4,
asymptotically de Sitter space regions are realized
in the neighborhood of the solutions.
 This is illustrated in Fig.3(a).
The oscillating curve (*) around the broken line
and the curve (**) represent
$
y=M -  {a^2 \omega_n ^2 \over 2\lambda}\, t
   - {a^2 \omega_n /\,4 \over \sqrt{\lambda^2 + 4\omega_n ^2}}
    \sin(2\omega_n t + \theta)\, \cos(2\omega_n x) \,
$
and
$ y =  - e^{-2\lambda t} $
respectively.
In order to have the degenerate solutions, the slope of the curve (*) must be
positive at some $t$. The first derivative of (*) is
$$
 {\partial y \over \partial t}
 = - {a^2 \omega_n ^2 \over 2\lambda}\bigl[ 1
    +{1  \over \sqrt{1 + (2\omega_n / \lambda)^2}}
    \cos(2\omega_n t + \theta )\, \cos(2\omega_n x) \bigr]\, .
\eqno\eq
$$

{}From this, it is obvious that the slope is monotonously down in any
region.
Thus we conclude that any degenerate solution cannot exist, that is,
 all the observer reach the singularity in the finite proper time inevitably.

On the other hand in the case of type-II, the solution has positive sign
in front of the second and the fourth terms
of the right hand side of (5.4):

$$
\eqalign{
 e^{-2\phi}
 =& M
   +  {a^2 \omega_n ^2 \over 2\lambda}\, t
   + e^{-2\lambda t}
   + {a^2 \omega_n /\,4 \over \sqrt{\lambda^2 + 4\omega_n ^2}}
    \sin(2\omega_n t + \theta)\, \cos(2\omega_n x) \, ,           \cr
 \tan (\theta)
 =& -{2\omega_n \over \lambda} \, .                               \cr
        }
\eqno\eq
$$

 In this case, we can control the parameters $M$ and $a$ so that
the degenerate solution $t_0$ occurs as shown in Fig.3(b).
Then we conclude that the observers whose world lines approach
the point $(t_0 , x_0)$
regard the point as
future infinity $\tilde{i}^+ =(t_0 , x_0)$.
In other words, such observers can avoid the singularity.

Before closing this section, we briefly comment on the three types
of global structure classified according to the behavior of
 the function $ g(t,x) \equiv e^{-2\phi}$
in the neighborhood of $(t_0 , x_0)$.
 For this purpose, $g(t,x)$ is supposed to be expanded around $(t_0 , x_0)$
as follows,
$$
\eqalign{
 g(t,x)
 =& g_x (t_0 , \, x_0) \,(x-x_0)
    +g_{tt}(t_0, \, x_0)\,(t-t_0)^2            \cr
  & +2g_{tx}(t_0 , \, x_0)\,(x-x_0)(t-t_0)
    +g_{xx}(t_0 \, x_0)\,(x-x_0)^2
    +... \  ,                                    \cr
          }
\eqno\eq
$$
where $g_{tt}(t_0, \, x_0)\, >\, 0 \, $.

According to the characteristics of the quadratic form in (5.7),
we can see the following behavior of the universe
in the neighborhood of  $\tilde{i}^+ =(t_0 , x_0)$ .

If
$
  g_x (t_0 , x_0) = 0 \,
$,
then the point   $ (t_0 , x_0) $ becomes
the local minimum
 or
the stationary point of  $y = g(t, x)$
(called case(1) and case(2) respectively), because
$  g_{tt}(t_0, \, x_0)\, >\, 0  $ .
In the former case (1), there is no singularity where the scalar
curvature diverges, as shown in Fig.4(1).
While in the latter case (2), there exists a singular line which approaches
to
 $\tilde{i}^+ =(t_0 , x_0)$ .
Whether the singular line is timelike or not (i.e. naked singularity appears
or not) depends on the form of the coefficient of (5.7).

If
$
 g_x(t_0 , x_0) \neq 0
$,
then case (3) arise, where the singular line is tangent to $x = x_0$.
This means that the naked singularity always emerges, as shown in
Fig.4(3).

For all the cases (1) $\sim$ (3), Cauchy horizon appears, shown as the
dashed-and-dotted lines.
These Cauchy horizons are regarded as `blue sheet' in general relativity.

\chapter{Including back-reaction}

In the second half of the paper, we study how the classical solution changes
if we include back-reaction.
For a while we treat only the model of type-I,
because the model of type-II can be treated in the similar manner.

In 1+1 dimensions, the quantum effect of massless matter fields is
completely determined by conformal anomaly[\polya].
The quantum effective action is sum of the classical action (2.1) and
the Polyakov term induced by the $N$ matter fields:
$$
 S_{quantum}
  = - {\kappa \over 8\pi} \int d^2x\sqrt{-g(x)}
                          \int d^2{\xchon}\sqrt{-g(\xchon )}
              R(x) \, G(x, \, \xchon) \,R(\xchon)  \> ,
\eqno\eq 
$$
where $\kappa$ is ${N\over 12}$ and  $\  G(x, \, \xchon)$ is a
Green's function of the scalar fields.
We assume that
$\kappa$ is to be
a large number and use the $1/N$-expansion.
Further we add the following term introduced
by Russo et al.[\rsti] to the above action:
$$
 S_{quantum}^\prime
 = - {\kappa\over 8\pi} \int d^2x\>\sqrt{-g} 2\phi \,R \> .
\eqno\eq 
$$

	Making the field redefinition:
$$
\eqalign{
  \chi
  =& \sqrt{\kappa}\bigl( \, \rho -{1 \over 2}\phi
                                 + {1 \over \kappa}e^{-2\phi}\,
                  \bigr)\, ,                                       \cr
  \Omega
  =& \sqrt{\kappa}\bigr( \, {1 \over 2}\phi +{1 \over \kappa}e^{-2\phi}\,
                  \bigr)\, ,                                       \cr
       }
\eqno\eq 
$$
the quantum equations of motion becomes in very simple
form.
The  effective action
$$
 S_{eff}
 = S_{cl}\, + \, S_{quantum}\, + \, S_{quantum}^\prime \, ,
\eqno\eq 
$$
is expressed in terms of
reference metric$\, \hat g_{\mu \nu}$
and the new fields $\chi$, $\Omega$ :

$$
\eqalign{
 S_{eff}
 =& {1\over \pi} \int d^2x\>\sqrt{-\hat g}\bigl[
    {1 \over 2} \hat g^{\mu \nu} \partial_{\mu}\chi \partial_{\nu}\chi
  - {1 \over 2} \hat g^{\mu \nu}
     \partial_{\mu}\Omega \partial_{\nu}\Omega                        \cr
  & \ \ \ \ - 2\lambda^2 e^{{2 \over \sqrt{\kappa}}(\chi - \Omega)}
  -{1 \over 4} \hat g^{\mu \nu} \partial_{\mu}f \cdot \partial_{\nu}f \cr
  & \ \ \ \ + {\sqrt{\kappa} \over 2} \chi \hat R
  -{\kappa \over 8} \hat R \hat \nabla^{-2} \cdot \hat R
                                          \bigr]\> \, .               \cr
       }
\eqno\eq 
$$
Then equations of motion in conformally flat gauge are
$$
\eqalign{
 &\partial_{+}\partial_{-} \chi
 = -{\lambda^2 \over \sqrt{\kappa}}
             \, e^{{2 \over \sqrt{\kappa}}(\chi - \Omega)} \, , \cr
 &\partial_{+}\partial_{-}( \chi - \Omega)
 = \, 0   \, ,                                                     \cr
        }
\eqno\eq
$$
and
$$
 \partial_{\pm}f \cdot \partial_{\pm}f -2\kappa t_{\pm}(x_{\pm})
 = -\partial_{\pm}\chi \partial_{\pm}\chi
   +  \partial_{\pm}\Omega \partial_{\pm}\Omega
   +\sqrt{\kappa} \partial^2 _{+}\chi \, ,
\eqno\eq
$$
where $\cdot$ denotes the sum over $i$, and the symbol $t_{\pm}$ is
arbitrary chiral
function to be determined by the boundary conditions. `the classical fields'
$f$ describe
macroscopic behavior of quantized fields.

Before proceeding the analysis, we determine $t_{\pm}$ in (6.7).
First, we recall that the vacuum expectation value (abbreviated as $vev$)
 of the stress tensor in the spacetime with the metric
$g_{\mu \nu}=e^{2\rho} \eta_{\mu \nu}$
is related to the $vev$ of the stress tensor in the spacetime with
$g_{\mu \nu}= \eta_{\mu \nu}$ [\bida]:
$$
 <\; T_{\pm\, \pm}^{f}[g_{\mu \nu}=e^{2\rho} \eta_{\mu \nu}]\;>
 = <\; T_{\pm\, \pm}^{f}[g_{\mu \nu}= \eta_{\mu \nu}]\; >
   -\kappa \bigl[\; (\partial_{\pm}\rho)^2  - \partial_{\pm} ^2 \rho \; \bigr].
\eqno\eq
$$
{}From this, we know that $-\kappa t_{\pm}$ represents
$<\, T_{\pm\, \pm}^{f}[g_{\mu \nu}= \eta_{\mu \nu}]\, >$ .
We consider the case of the closed universe so that
$<\, T_{\pm\, \pm}^{f}[g_{\mu \nu}= \eta_{\mu \nu}]\, >$
 is set to the one in the static cylindrical universe.
In this case, due to the Casimir effect,
$ \, <\, T_{\pm\, \pm}^{f}[g_{\mu \nu}= \eta_{\mu \nu}]\, >$
becomes
$-{\kappa \pi ^2 \over 12 L^2} \, $ .
Thus we obtain $t_{\pm}$ as
$$
t_{\pm}
= {\pi ^2 \over L^2}  \, .
\eqno\eq
$$

\chapter{Quantum behavior of the universe}

 For the quantum state with an arbitrary macroscopic part of quantized
matter fields as in (5.1), the corresponding solution of
the equations (6.6) and (6.7) is given by
$$
\eqalign{
 \sqrt{\kappa} \chi
  =& M - {2\kappa \over \lambda}
    \bigl( \, {\lambda^2 \over 4} - t_{vev}
    + {\varepsilon + \Sigma \over 4\kappa} \bigr) \, t
    + e^{-2\lambda t}
    + \bigl(\, oscillation\  part \, \bigr)\, ,   \cr
 \sqrt{\kappa} \Omega
  =& M + {2\kappa \over \lambda}
    \bigl( \, {\lambda^2 \over 4} + t_{vev}
    - {\varepsilon + \Sigma \over 4\kappa} \bigr) \, t
    + e^{-2\lambda t}
    + \bigl(\, oscillation\  part \, \bigr)\, ,                       \cr
        }
\eqno\eq
$$
where $t_{vev}=t_{\pm}={\pi ^2 \over L^2}$ determined in the previous section,
and $\Sigma$ is the same as in (5.2).
{}From the solution (7.1) and the field redefinition (6.3),
we confirm that the behavior of $\rho$ and $\phi$
in the distant past agree with the classical one, as expected,
because
$ \sqrt{\kappa} \Omega $
approaches to
$e^{-2\phi}$
as $t \rightarrow - \infty$ .
It should be also noted that
another new singularity appears if the universe takes
the value $\phi = \phi_{cr}$ corresponding
to local minimum of $\Omega$ (see Fig.5(a)).

 The behavior of the universe is classified into three types as follows.
In the case: $\Omega_{cr} > \Omega_{min}$ (see Fig.5(a,b)),
the universe starts at the weak-coupling region, and as $t$ goes on,
the $\phi$ grows gradually,
then returns to the weak-coupling region before $\Omega$ becomes $\Omega_{cr}$.
 In the case: $\Omega_{cr} < \Omega_{min}$, the universe reaches the point at
$\phi = \phi_{cr}$ and end the evolution.
The case: $\Omega_{cr} = \Omega_{min}$ is rather subtle.
More detailed analysis clarifies that the spacetime is regular
at $\phi = \phi_{cr}$ so that the universe can evolve into
the strong coupling region.

 Next we consider whether a quantum analogue to each classical behavior
stated before exists.

 In the first, the static cylindrical universe is considered.
{}From (6.5) and (7.1), we obtain the relation,
$$
 2\rho
 = 2\phi + {2 \over \sqrt{\kappa}}(\, \chi - \Omega \,) .
\eqno\eq
$$
Then we easily recognize that
the static cylindrical universe is realized if
$\rho = 0$ (i.e. $2\phi = 2\lambda t $).
In this case, the parameters $M$ and $\varepsilon$ must be adjusted as
$$
\eqalign{
M\;
 =& \; 0\, ,               \cr
 \varepsilon \;
 =& \; 4\kappa t_{vev} \, .    \cr
         }
\eqno\eq
$$

 In the second, to extend the spacetime
to the region with CTC's, the following two conditions are necessary and
sufficient in the conformal gauge: One is that $2 \rho $ become linear in $t$
as $t \rightarrow \infty$ in the conformal flat gauge,
and the other, the coefficient of the linear term is negative.
By comparing $\Omega \, - \, \phi$ and $\Omega \, - \, t$ relations,
we recognize that there are two distinct types of solutions.
 One is realized in the case (i): $ M >\, \sqrt{\kappa} \, \Omega_{cr}\,$
with the parameters
$$
\eqalign{
&a^{(\pm)}_{n}
 =\, b^{(\pm)}_{n} = 0 \ \ \ \ \ \ \ \ \     \forall{n}\, ,  \cr
& \varepsilon
 =\, \kappa \lambda^2 + 4\kappa t_{vev}\, .
       }
\eqno\eq
$$
The first condition in (7.4) means that the universe is homogeneous.
The other case is realized in the case (ii): $ \Omega_{min} = \Omega_{cr} $,
$ a^{(\pm)}_{n} =\, b^{(\pm)}_{n} = 0 \, (\forall{n})\, $ and some appropriate
negative $M$. In the case (i), the value of dilaton at $\eta = 0$
can be adjusted to be so small that the semi-classical approximation is valid.
On the other hand in the case (ii), the spacetime can be extended over
$ \phi = \phi_{cr} $ to the strong coupling region smoothly and
the appearance of CTC-spacetime may occur in the limit of
$t = \infty$ where $\phi$ becomes infinite.
Hence the analysis with full quantization is needed so that the statement that
whether CTC's appear may becomes meaningless in the case (ii).

 In the third, we consider the possibility of the local or global
de Sitter expansion briefly.
Following the discussion in the last part of the section 4,
de Sitter expansion occurs only
when $e^{2\phi}$
 behave as $(t-t_0)^{-2}$ around at some finite time $t_0$.
We easily know that such a case can never exist
owing to the existence of the linear term with respect to
$\phi$ in (6.3).

 In the rest of this section, we treat the case of type-II.
The analysis can be proceeded in the same way as type-I.
The effective action in type-II is also described,
using new fields $\chi$ and $\Omega$:
$$
\eqalign{
 S_{eff}
 =& {1\over \pi} \int d^2x\>\sqrt{-\hat g}\bigl[
    {1 \over 2} \hat g^{\mu \nu} \partial_{\mu}\chi \partial_{\nu}\chi
  - {1 \over 2} \hat g^{\mu \nu}
     \partial_{\mu}\Omega \partial_{\nu}\Omega                        \cr
  & \ \ \ \ + 2\lambda^2 e^{{2 \over \sqrt{\kappa}}(\chi - \Omega)}
  -{1 \over 4} \hat g^{\mu \nu} \partial_{\mu}f \cdot \partial_{\nu}f \cr
  & \ \ \ \ + {\sqrt{\kappa} \over 2} \chi \hat R
  -{\kappa \over 8} \hat R \hat \nabla^{-2} \cdot \hat R
                                          \bigr]\> \, ,               \cr
       }
\eqno\eq
$$
where
$$
\eqalign{
  \chi
  =& \sqrt{\kappa}\bigl( \, \rho -{1 \over 2}\phi
                                 - {1 \over \kappa}e^{-2\phi}\,
                  \bigr)\, ,                                       \cr
  \Omega
  =& \sqrt{\kappa}\bigr( \, {1 \over 2}\phi - {1 \over \kappa}e^{-2\phi}\,
                  \bigr)\, .                                       \cr
       }
\eqno\eq
$$
The equations of motion derived from (7.5) is
$$
\eqalign{
 \partial_{+}\partial_{-} \chi
 =& {\lambda^2 \over \sqrt{\kappa}}
             \, e^{{2 \over \sqrt{\kappa}}(\chi - \Omega)} \, , \cr
 \partial_{+}\partial_{-}( \chi - \Omega)
 =& \, 0 \, .                                                     \cr
        }
\eqno\eq
$$
The constraint equation is as same as (6.7).

 The solution for $\Omega$ is
$$
\eqalign{
   - \sqrt{\kappa} \Omega
  =& M + {2\kappa \over \lambda}
    \bigl( \, {\lambda^2 \over 4} - t_{vev}
    + {\varepsilon \over 4\kappa} \bigr) \, t
    + e^{-2\lambda t}  \, .                                  \cr
        }
\eqno\eq
$$

 We can now investigate the behavior of the universe of type-II
in the same way as type-I.
In contrast to type-I,
$\Omega \, -  \phi$ graph is not bounded from below
from the second equation (7.6),
that is, $\phi_{cr}$ (i.e. the point that quantum singularity emerges)
does not exist.

 There are three types of behavior of the universe: Firstly, in the case:
$t_{vev} < {\lambda^2 \over 4} + {\varepsilon \over 4\kappa}$,
the universe returns to the weak-coupling region. Secondly, in the case:
$t_{vev} > {\lambda^2 \over 4} + {\varepsilon \over 4\kappa}$,
the universe evolve into the strong-coupling region.
Thirdly, in the case:
$t_{vev} = {\lambda^2 \over 4} + {\varepsilon \over 4\kappa}$,
the universe becomes the Misner-type universe at $\phi = \phi_{min}$
as $t$ goes to $+ \infty$.

Finally, we note that any de Sitter-like behavior of the universe never
appear according to the same reason as mentioned in type-I.

\chapter{Conclusion}

 We have applied CGHS model to 1+1 dimensional cosmology.
In the classical level, we classified the behavior of
the universe evolving from a static cylindrical one.
 Though some solutions derived in this paper have unusual behavior,
we have some interesting spacetime
which would give us new insights about the global structure
of our real four-dimensional universe. Next their modifications
by considering inhomogeneous matters or quantum back-reaction
were considered.
 Especially, instability of asymptotically de Sitter universe and
Misner-type universe have been treated.
Firstly, due to inhomogeneity of matters,
asymptotically de Sitter universe is modified
into the universe having partially de Sitter-like expansion regions
(two-dimensional analogue of multi-production of universe),
on the other hand, Misner-type universe is completely forbidden to appear.
 Secondly, including back-reaction, we showed that
an asymptotically de Sitter region cannot exist any longer,
and Misner-type universe still
exists in the case with adjusted parameters stated in the
previous section.

 In order to determine exactly whether the chronology protection holds,
we must go on to extend the classical and semi-classical treatments
of the compact universe in this paper
to the full quantization of two-dimensional dilaton
gravity (for example, see [\mntu, \kaz]).
Because the existence of any consistent
solution in the extended region ($\eta > 0$) inevitably depends on
the information from the naked singularity($\eta = M$), whose neighborhood
is very strong coupling region. To proceed the analysis, the
construction of the physical states having such classical
and semi-classical behaviors
will be intriguing especially.

 Anyhow CGHS scenario has been shown to be extensively useful for the
study of back-reaction problems in general relativity.
\vskip 1 cm

\subsection{Acknowledgment}
 We are most grateful to A. Hosoya for a collaboration in an early stage of the
work. One of the authors(T. M.) would like to acknowledge Y.~Onozawa, M.~Siino
and K.~Watanabe for enjoyable discussions.

\refout
\vskip 2cm


\noindent {\bf Figure Captions}

\vskip 1.5cm
\item{\bf Fig. 1}Classification of the universe with respect to the
                 sign of $M$ ;
\nextline
(a) $M = 0,$  (b) $M > 0,$  (c) $M < 0$ .
\vskip 1cm

\item{\bf Fig. 2}Misner-type universe;
\nextline
$t=+\infty \, (\eta=0)$ is a closed null geodesics. The region:$(\eta<0)$ is
globally hyperbolic spacetime, and the region:$(\eta>0)$ have closed
time-like curves.
\vskip 1cm

\item{\bf Fig. 3}Behavior of the solution of $e^{-2\phi} = 0$ ;
\nextline
The curve (*) around the broken line
and the curve (**) represent
$
y=M -  {a^2 \omega_n ^2 \over 2\lambda}\, t
   - {a^2 \omega_n /\,4 \over \sqrt{\lambda^2 + 4\omega_n ^2}}
    \sin(2\omega_n t + \theta)\, \cos(2\omega_n x) \,
$
and
$ y =  - e^{-2\lambda t} $
respectively.
\nextline
The cases of type-I and type-II have degenerate solutions
 in graphs (a) and (b) respectively.
\nextline
\vskip 1 cm

\item{\bf Fig. 4}Possible configurations of time-like infinity
$\tilde{i}^+$ and singularity;
\nextline
The symbol $\times$ denotes the time-like infinity $\tilde{i}^+$ ,
broken line denotes the singularity,
dashed-and-dotted line denotes Cauchy Horizon from the
time-like infinity $\tilde{i}^+$.

\item{\bf Fig.5}$\Omega \,$-$\, \phi$ and $\Omega \,$-$\, t$ relations;
\nextline
(a) $\Omega \,$-$\, \phi$ relation, (b) $\Omega \,$-$\, t$ relation
\nextline
$\Omega_{cr}$ is local minimum of the
function $\Omega$, with respect to $\phi$.
$\Omega_{mim}$ is local minimum of the function $\Omega$,
with respect to t.

\bye